\documentclass[epj,amssymb,floatfix,superscriptaddress,preprintnumbers]{svjour}
\usepackage{graphics}
\usepackage[utf8]{inputenc}
\usepackage[T1]{fontenc}
\usepackage{bm}
\usepackage{float}
\usepackage[usenames]{color}
\usepackage[normalem]{ulem}
\usepackage{graphicx}
\usepackage{amsmath}
\usepackage{amsfonts}
\usepackage{dsfont}  
\usepackage{slashed}  
\usepackage{booktabs}
\usepackage{multirow}
\usepackage{subfigure}
\DeclareUnicodeCharacter{2212}{-}

\begin{document}
%
%
\title{Charge radii of the nucleon from its flavor dependent Dirac form factors}
\author{H. Atac\inst{1} \and M. Constantinou \inst{1} \and Z.-E. Meziani \inst{2,1} \and M. Paolone \inst{3,1} \and N. Sparveris \inst{1}
\thanks{Corresponding author e-mail: sparveri@temple.edu}%
}                     
%
%
\institute{Temple University, Philadelphia, PA 19122, USA \and Argonne National Laboratory, Lemont, IL 60439, USA \and New Mexico State University, Las Cruces, NM 88003, USA}
\date{Received: date / Revised version: date}
%
\abstract{
We have determined the proton and the neutron charge radii from a global analysis of the proton and the neutron elastic form factors, after first performing a flavor decomposition of these form factors under charge symmetry in the light cone frame formulation. We then extracted the transverse mean-square radii of the flavor dependent quark distributions. In turn, these are related in a model-independent way to the proton and neutron charge radii but allow us to take into account motion effects of the recoiling nucleon for data at finite but high momentum transfer.
In the proton case we find $\langle r_p \rangle = 0.852 \pm0.002_{\rm (stat.)} \pm0.009_{\rm (syst.)}~({\rm fm})$, consistent with the proton charge radius obtained from muonic hydrogen spectroscopy~\cite{pohl:2010,antog2013}. The current method improves on the precision of the $\langle r_p \rangle$ extraction based on the form factor measurements.
Furthermore, we find no discrepancy in the $\langle r_p \rangle$ determination among the different electron scattering measurements, all of which, utilizing the current method of extraction, result in a value that is consistent with the smallest $\langle r_p \rangle$ extraction from the electron scattering measurements~\cite{Xiong:2019umf}.
Concerning the neutron case, past results relied solely on the neutron-electron scattering length measurements, which suffer from an underestimation of underlying systematic uncertainties inherent to the extraction technique. Utilizing the present method we have performed the first extraction of the neutron charge radius based on nucleon form factor data, and we find $\langle r_n^2 \rangle = -0.122 \pm0.004_{\rm (stat.)} \pm0.010_{\rm (syst.)}~({\rm fm}^2)$.
}

\maketitle
\section{Introduction}
\label{intro}
The study of the nucleon charge radius has been historically instrumental towards the understanding of the nucleon structure. 
For the proton, in atomic spectroscopy, the sensitivity of its size to the atomic energy levels is determined by the probability that the bound lepton be within the volume of the proton.
This probability is approximately given by the ratio of proton to atomic volumes. The muon thus offers higher sensitivity in the determination of the proton's charge radius since it is about 8 million more times likely to be inside the proton than the electron, a consequence of the muon mass being about 200 times the mass of the electron. Electron scattering has also long been utilized for the measurement of the proton charge radius. In this case the radius is determined by the slope of the electric form factor of the proton at four-momentum transfer $Q^2=0$. 
A significant challenge here lies with the choice of the adopted functional forms that are fitted and with the quantification of the resulting model uncertainties.
The disagreement of the proton charge radius, $\langle r_p\rangle$, as determined using the measurement of the Lamb shift in the muonic hydrogen atom~\cite{pohl:2010}, with the earlier results based on the hydrogen atom and the electron scattering measurements gave rise to the proton radius puzzle~\cite{protonpuzz:2013,bernpohl:2014}. This, in turn, led to a significant reassessment of the methods and analyses utilized in the radius extraction, as well as to the consideration of physics beyond the standard model, as potential solutions to this discrepancy. Recent tensions between the spectroscopic measurements conducted on hydrogen~\cite{Flaurbaey:2018,bezginov:2019} and between the electron scattering measurements~\cite{Xiong:2019umf} have further complicated this puzzle.
In the case of the neutron it is the highly complicated dynamics of the strong force between quarks and gluons that lead to an asymmetric distribution of the u- and d-quarks in the system, resulting in a negative value for $\langle r_n^2 \rangle$. Therefore, the precise measurement of $\langle r_n^2 \rangle$ also represents a critical part in our understanding of the nucleon dynamics. In contrast to the proton case, the $\langle r_n^2\rangle$ determination is more challenging since no equivalent atomic technique is possible, and the electron scattering method suffers from severe limitations due to the absence of a free neutron target. 
Thus, the $\langle r_n^2 \rangle $  extraction has been solely based on the measurement of the {\it neutron}-electron scattering length where low-energy neutrons are scattered by electrons bound in diamagnetic atoms. 
The $\langle r_n^2 \rangle$ measurements adopted by the particle data group (PDG)~\cite{Kopecky:1997rw,Koester:1995nx,Aleksandrov:1986mw,Krohn:1973re}, the most recent of which is dated two decades ago, exhibit discrepancies with values ranging from $\langle r_n^2 \rangle = -0.115 \pm0.002 \pm0.003~({\rm fm}^2)$~\cite{Kopecky:1997rw} to $\langle r_n^2 \rangle = -0.134 \pm0.009~({\rm fm}^2)$~\cite{Aleksandrov:1986mw}. Among the plausible explanations suggested are the effect of resonance corrections and of the electric polarizability, as discussed e.g. in~\cite{Koester:1995nx}. 
However, these discrepancies have not been fully resolved, which reveals the limitations of this method and indicates a potential underestimation of the underlying systematic uncertainties.
Considering the fundamental symmetry between the two isospin partners, it becomes evident that being able to employ alternative methods in extracting $\langle r^2_n \rangle$ may prove most valuable, as recently exhibited in the proton's case. 
In this work we perform first a flavor decomposition of the Dirac form factors using the proton and neutron Sachs form factor world data and show how well they compare to the most up-to-date lattice calculations. Next, using the light cone frame formulation of these form factors, the transverse mean-square radii of the up and down quarks are extracted. 
Through model-independent relations the mean-square radii of the proton and neutron are determined and their stability with respect to form factors fitting models and the $Q^2$ range is explored. Values of the proton and neutron radii are determined and their validity is discussed.

\begin{figure}
\centering
\resizebox{\linewidth}{!}{%
\includegraphics{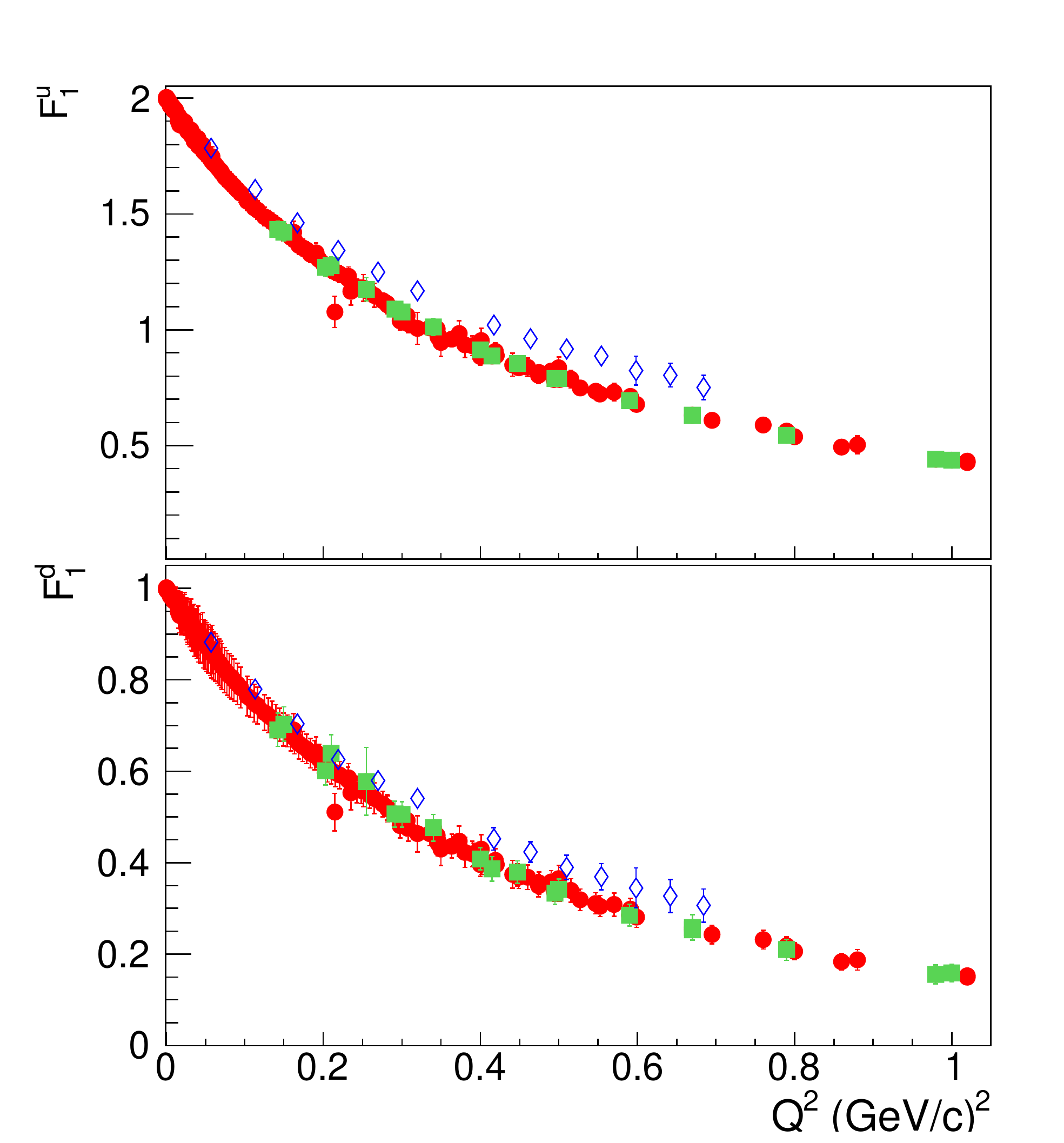}
}
\caption{The $F_{1}^{u}$ (top panel) and the $F_{1}^{d}$ (bottom panel) as extracted from the nucleon form factor world data. The filled (red) circles mark the experiments of the proton form factors \cite{Bernauer.2010.105.242001,Xiong:2019umf,Bernauer:2013tpr,Punjabi:2005wq,Gayou.64.038202,strauch.91.052301,milbrath.82.2221,Pospischil:2001pp,crawford.98.052301,Ron.99.202002,Zhan:2011ji,Paolone:2010qc} and the filled (green) boxes those of the neutron form factors~\cite{Geis:2008aa,Herberg:1999ud,Passchier:1999ju,Eden.50.R1749,Glazier:2004ny,ostrick.83.276,golak.63.034006,Madey:2003av,Zhu:2001md,Warren:2003ma,Rohe:1999sh,Bermuth:2003qh,sulko}. The lattice results~\cite{Alexandrou:2018sjm} are shown with open (blue) diamonds.}
\label{fig1}
\end{figure}

\section{Flavor decomposition of the Dirac form factors}
\label{sec:1}
We perform the flavor decomposition of the elastic nucleon electromagnetic form factors~\cite{npn:2011} by conducting an analysis combining the proton and the neutron Sachs form factor world data, namely~\cite{Bernauer.2010.105.242001,Xiong:2019umf,Bernauer:2013tpr,Punjabi:2005wq,Gayou.64.038202,strauch.91.052301,milbrath.82.2221,Pospischil:2001pp,crawford.98.052301,Ron.99.202002,Zhan:2011ji,Paolone:2010qc,Geis:2008aa,Herberg:1999ud,Passchier:1999ju,Eden.50.R1749,Glazier:2004ny,ostrick.83.276,golak.63.034006,Madey:2003av,Zhu:2001md,Warren:2003ma,Rohe:1999sh,Bermuth:2003qh,sulko}. For the proton we have considered the recent high precision results from the two experiments that exhibit tension to the extraction of $\langle r_p\rangle$, namely the ~\cite{Xiong:2019umf} and~\cite{Bernauer.2010.105.242001}, as well as the polarization data, but we have not included older cross section measurements. Here we follow the same line of work as previously done with the high-$Q^2$ measurements in~\cite{Cates:2011pz}. The Dirac form factor, $F_1$, is expressed in terms of the Sachs form factors through 
\begin{equation}
\label{F1}
 F_1 = (G_E+\tau G_M) / (1+\tau)\,,
\end{equation}
where $\tau=Q^2/4m_N^2$ and $m_N$ the mass of the nucleon. Utilizing the Dirac nucleon form factors, $F_1^{\it p(n)}$ we perform the flavor decomposition of the form factors under charge symmetry using the relations
\begin{eqnarray}\label{flavdecomp} 
F_{1}^u \, = \, 2\,F_{1}^p \,+\, F_{1}^n \\[1ex]
F_{1}^d \, = \, 2\,F_{1}^n \,+\, F_{1}^p
\end{eqnarray}
where with $F_{1}^u$ and $F_{1}^d$ we refer to the up and down quark contributions to the Dirac form factors of the proton.
The normalizations of the Dirac form factors at $Q^2=0$ are given by $F_{1}^{u}(0) = 2$ and $F_{1}^{d}(0) = 1$ 
so as to yield a normalization of 2 and 1 for the $u$ and $d$-quark distributions in the proton, respectively.

\begin{figure*}
\centering

\includegraphics[width=\textwidth]{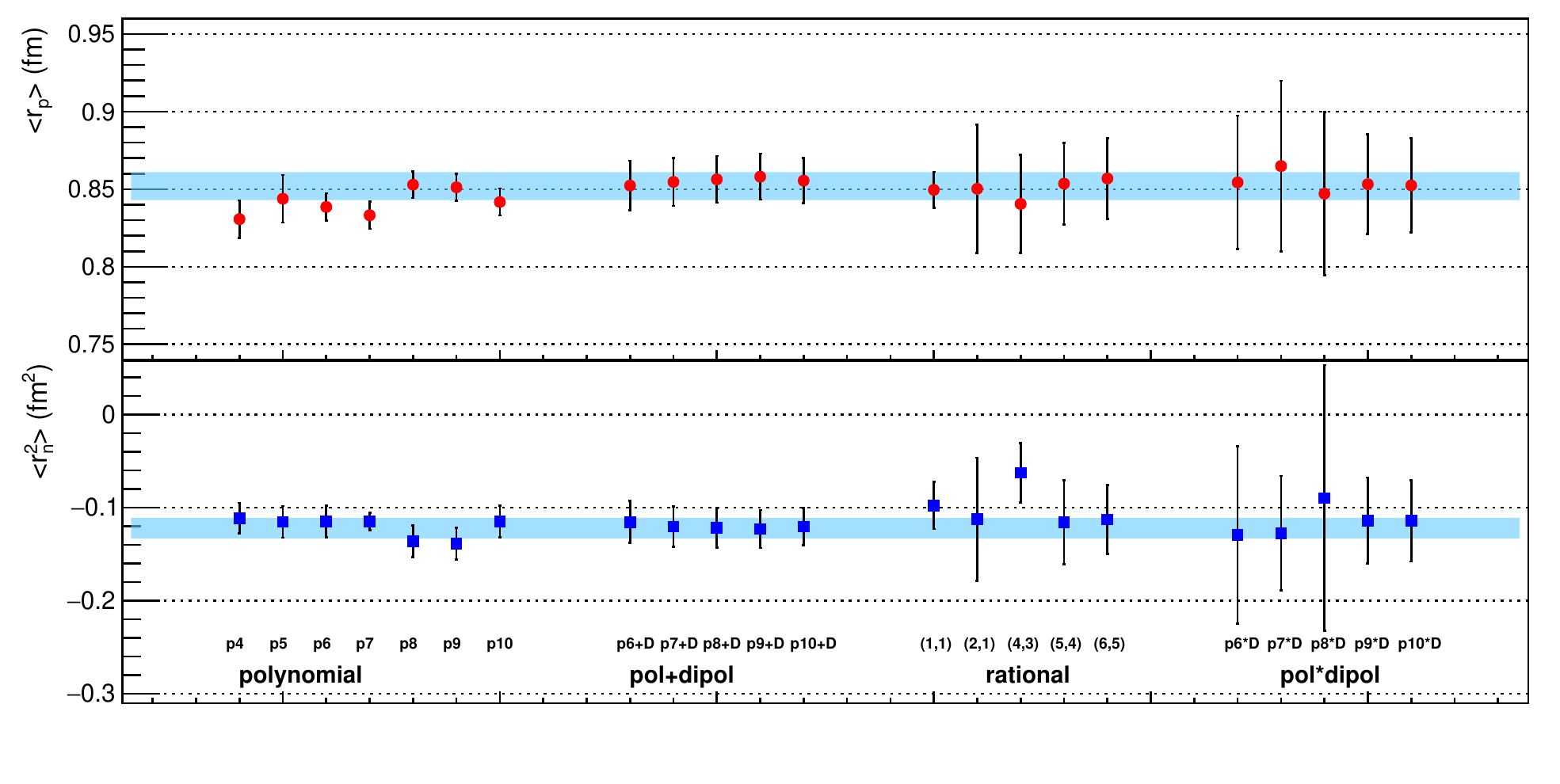}
\caption{The proton and neutron charge radius (top and bottom, respectively) as extracted from the fitted functions in all four groups, for a fitting range of $Q^2=[0,1]~(GeV/c)^2$. The error bars indicate the total (statistical+systematic) uncertainty of each fit.}
\label{fig7rev}
\end{figure*}

The proton and the neutron Sachs form factor world-data come from different measurements, and as such, they are not matched in $Q^2$. Thus, the Sachs form factor measurements for each of the proton and neutron world data are analyzed using a parametrization for its isospin partner counterpart, with a corresponding uncertainty that is propagated to the extracted flavor dependent form factors at each $Q^2$, so that the flavor decomposition of the Dirac form factors is performed. The analysis is then repeated with multiple parametrizations and the variance of the extracted results is accounted for as an additional uncertainty. A recent set of parametrizations that we utilize involve those presented in~\cite{Ye:2017gyb}. In order to further explore the effect of the choice of parametrization in the analysis, we have also derived updated parametrizations utilizing the widely used functional forms of~\cite{Kelly:2004hm} where we have included the most recent world data for the nucleon Sachs form factors, and we have considered these parametrizations in the analysis. Explicitly for the neutron electric form factor, we have considered the two parametrizations that are known to describe the world data, i.e. the Galster~\cite{Galster:1971kv} and the sum of two dipoles. So as not to bias the analysis, we have not adopted any constraints for the $G_E^n$ slope at $Q^2=0$ that are based on the neutron-electron scattering length extraction of $\langle r_n^2 \rangle$, something that was frequently applied to $G_E^n$ parametrizations in the past, and we have not used the $G_E^n$ parametrization in~\cite{Ye:2017gyb} that uses the neutron-electron scattering length measurement of $\langle r_n^2 \rangle$ as a data point in the fit. Instead, we introduce an additional (third) free parameter in both parametrizations and we allow the $\langle r_n^2 \rangle$ to remain unconstrained. For example, in the case of the Galster parametrization, instead of using the standard dipole form factor with $\Lambda^2=0.71({\rm GeV/c})^2$ we introduce an additional free parameter, namely
\begin{equation}\label{modgalster} G_E^n(Q^2)=(1 + Q^2/A)^{-2} \frac{B \tau}{1 + C \tau} , \end{equation}
where $A, B, C$ are free parameters.
The parameters obtained are $A = 0.542 \pm 0.155$, $B = 1.571 \pm 0.245$, $C = 1.075 \pm 0.947$. For the two dipole parametrization
\begin{equation}\label{2dipolfit} G_E^{n}(Q^2)= \frac{A}{(1+\frac{Q^{2}}{B})^2} -\frac{A}{(1+\frac{Q^{2}}{C})^2} \end{equation}
the parameters are $A = 0.138 \pm 0.071$, $B = 1.738 \pm 0.579$, $C = 0.449 \pm 0.193$.
The derived results for the flavor dependent Dirac form factors, $F_{1}^u$ and $F_{1}^d$, are shown in Fig.~\ref{fig1}. 
The results that utilize the measurements of the proton form factors \cite{Bernauer.2010.105.242001,Xiong:2019umf,Bernauer:2013tpr,Punjabi:2005wq,Gayou.64.038202,strauch.91.052301,milbrath.82.2221,Pospischil:2001pp,crawford.98.052301,Ron.99.202002,Zhan:2011ji,Paolone:2010qc} and a parametrization for the neutron form factors are shown with circles, while the boxes indicate those that utilize the neutron form factor measurements~\cite{Geis:2008aa,Herberg:1999ud,Passchier:1999ju,Eden.50.R1749,Glazier:2004ny,ostrick.83.276,golak.63.034006,Madey:2003av,Zhu:2001md,Warren:2003ma,Rohe:1999sh,Bermuth:2003qh} and a parametrization for the proton form factors.

The experimental results are compared to results from lattice QCD using twisted-mass fermions~\cite{Alexandrou:2018sjm}. $F_1$ has been extracted using results on the Sachs form factors $G_E^{p(n)}$ and $G_M^{p(n)}$ and Eq.~(\ref{F1}). The flavor decomposition of the $u$ and $d$ quark contributions is feasible because the lattice calculation for the proton and the neutron electromagnetic form factors include both the connected and disconnected diagrams. Also, the lattice calculation is performed at the physical value of the pion mass, eliminating a major source of systematic uncertainty; the chiral extrapolation. As can be seen in Fig.~\ref{fig1}, the lattice results exhibit a remarkable agreement with the experimental world data especially at small $Q^2$.

\begin{figure}
\centering
\resizebox{\linewidth}{!}{%
\includegraphics{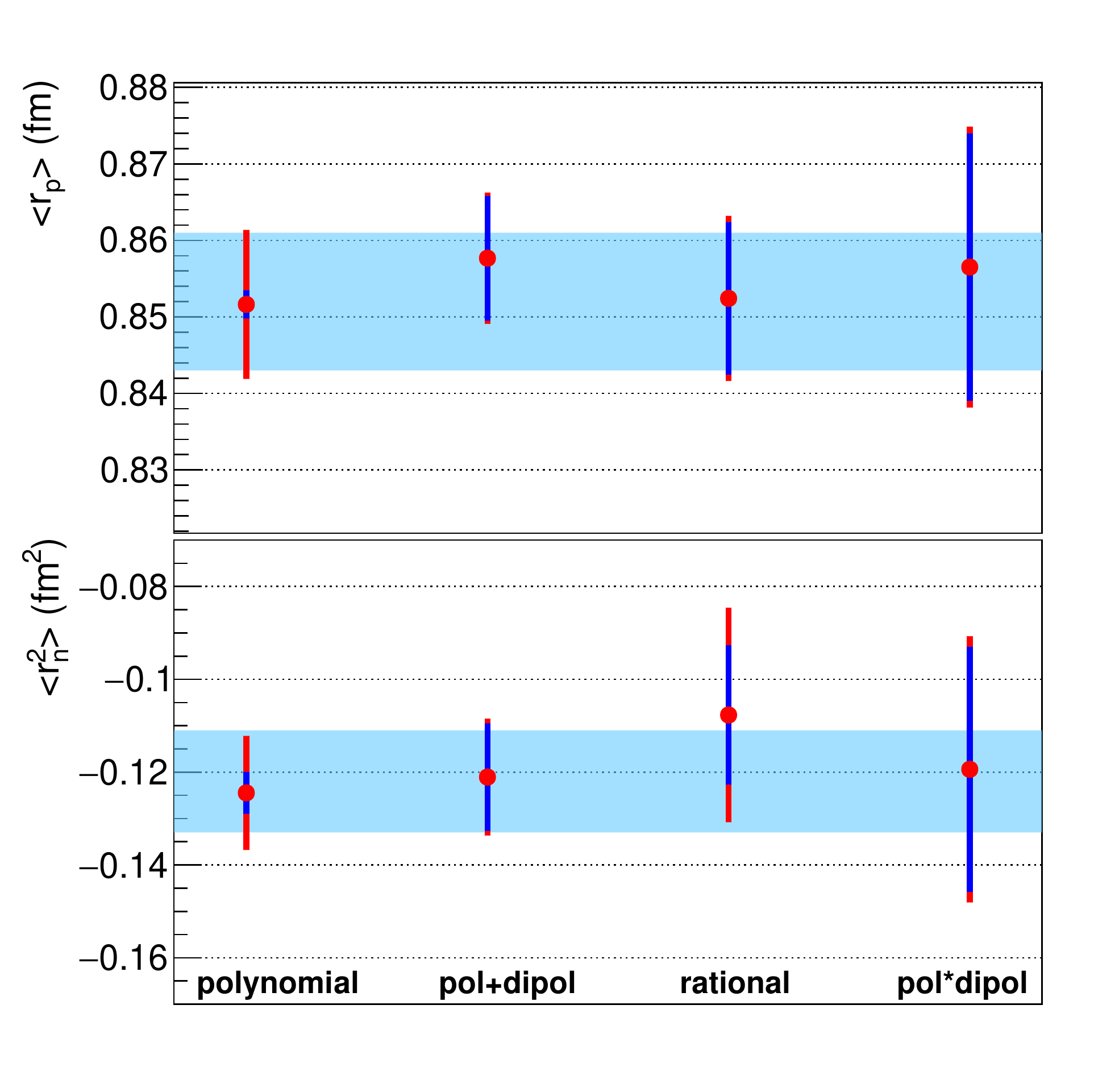}
}
\caption{The proton and neutron charge radius (top and bottom, respectively) as extracted utilizing the various groups of the fitted functions. The error bars correspond to the statistical and the total uncertainty, respectively. The polynomial statistical uncertainty for $\langle r_p \rangle$ is very small and the total uncertainty is effectively dominated by the systematic uncertainties in this case. The blue band marks the final result when all groups of functions are considered.}
\label{fig2}
\end{figure}

\section{Mean square radii dependence on fitting model and $Q^2$ range}
\label{sec:2}
The slopes of the flavor-dependent Dirac form factors at $Q^2=0$ are related to the  transverse mean-square radii of the 2-dimensional transverse quark charge distributions
\begin{equation}\label{slopequark} 
\langle b_{u(d)}^2 \rangle = \left. \frac{-4}{F_1^{u(d)}(0)}\frac{dF_1^{u(d)}(Q^2)}{dQ^2} \right\rvert_{Q^2 \rightarrow 0} 
\end{equation}
where {\it b} denotes the quark position in the plane transverse to the longitudinal momentum of a fast moving nucleon \cite{Miller:2007uy,Miller:2019,dupre:2017}. Once the $\langle b_{u(d)}^2 \rangle$ are determined from fits to the  $F_1^{u(d)}$ data (Fig.~\ref{fig1}), the proton and the neutron charge radii are extracted  through the model-independent relations
\begin{eqnarray}\label{protonradius}
 \langle r_p^2 \rangle &=&  2 \langle b_u^2 \rangle -\frac{1}{2} \langle b_d^2 \rangle + \frac{3}{2} \frac{\kappa_N}{M_N^2}
\\
 \langle r_n^2 \rangle &=&  \langle b_d^2 \rangle - \langle b_u^2 \rangle + \frac{3}{2} \frac{\kappa_N}{M_N^2}
\end{eqnarray}
where $\kappa_N$ and $M_N$ the anomalous magnetic moment and the mass of the nucleon.

The 3D Breit frame and 2D light-front distributions are directly related to each other~\cite{lorce} and the extraction of the nucleon charge radius through the transverse quark distributions is fundamentally equivalent to its determination through the slope of the nucleon's electric Sachs form factor at $Q^2=0$. However, extracting the radius can only be performed within finite and extended $Q^2$ range, and this fact introduces some sensitivity to all the nucleon form factors as the $Q^2$ increases. Nevertheless, unlike in the Breit frame, the infinite-momentum frame offers the inherent advantage that a true transverse charge density can be properly defined as the matrix element of a density operator between identical initial and final states mitigating relativistic nucleon recoil effects. An additional advantage emerges from the fact that the experimental data base expands to include the measurements of both the proton and neutron.

\begin{figure}
\centering
\resizebox{\linewidth}{!}{%
\includegraphics{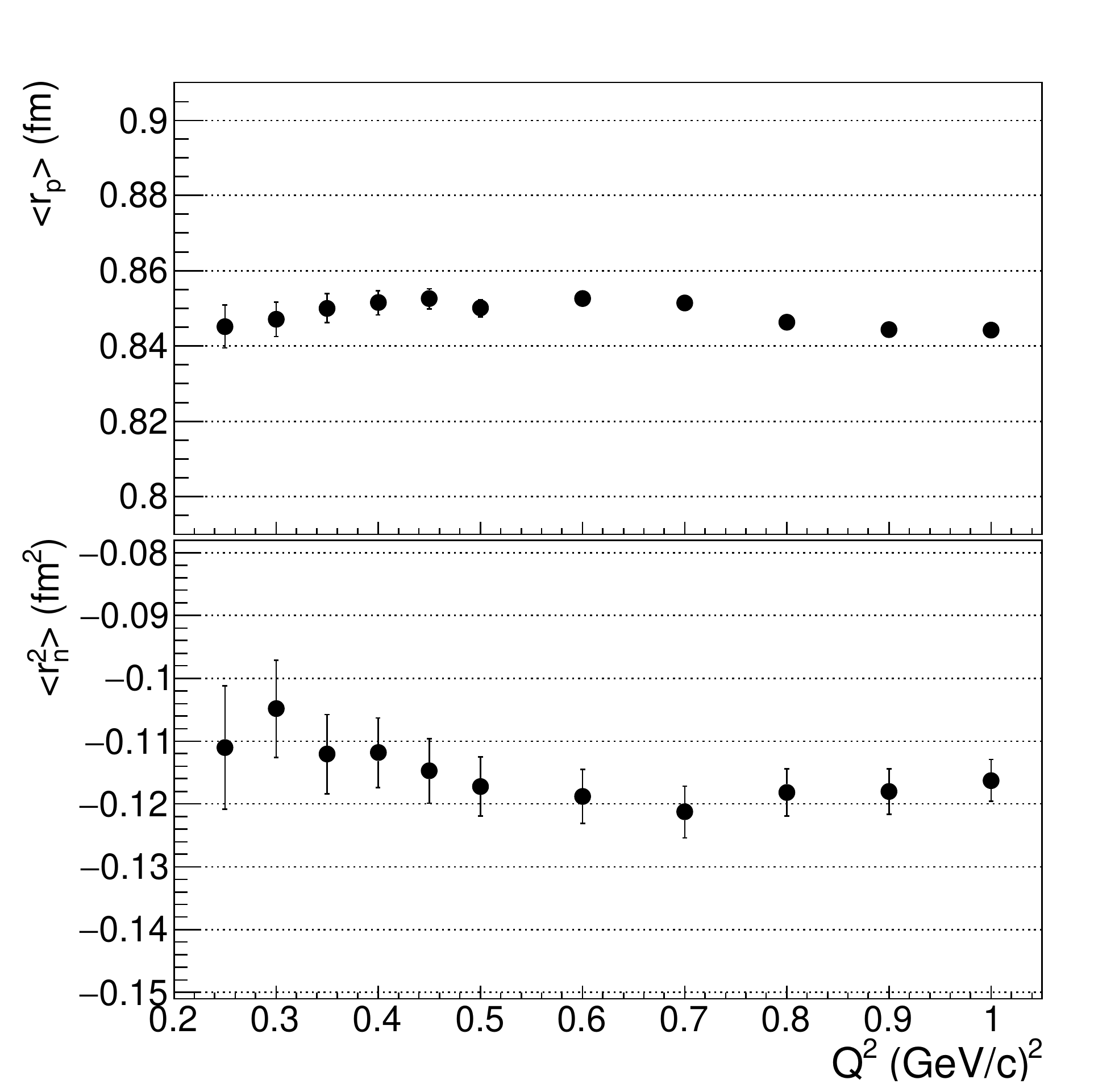}
}
\caption{The proton and neutron charge radius (top and bottom panels, respectively) resulting from the combination of all the groups of fitted functions as a function of the fitted range of momentum transfer; the $Q^2$ marks the upper bound of the fitting range. The error bars indicate the statistical uncertainty of the fits.}
\label{fig3}
\end{figure}

The fits are performed within the range of $Q^2=0$ to $Q^2=1~(GeV/c)^2$. The analysis stays within the kinematic range where the competing $\langle r_p \rangle$ extractions from the recent electron scattering measurements have been performed, namely from the PRad~\cite{Xiong:2019umf} and the MAMI~\cite{Bernauer.2010.105.242001} experiments. Furthermore, we do not extend the fits higher than $Q^2=1~(GeV/c)^2$ so as to avoid introducing any potential model dependence to the fits that is associated with the scaling, considering that the $F_1^u$ and the $F_1^d$ have to be both fitted simultaneously by the same functional form.
In order to determine the slope of $F_1^{u(d)}$ at $Q^2=0$ a variety of functional forms have been employed to fit the data, namely polynomial, polynomial+dipole, polynomial~$\times$~dipole, and rational functions of the form $$f(Q^2)=\frac{\alpha_0 + \sum\limits_{i=1}^{n} \alpha_i Q^{2i}}{1 + \sum\limits_{j=1}^{m} \beta_j Q^{2j}}.$$  These are the typical groups of functions that have been utilized in the past for the proton charge radius extraction.
We explored other functional forms, but have not included them as they were not able to offer a good fit. 
In order to investigate the stability of the extracted charge radii we used polynomials of different degree. We find that, polynomials above the 3$^{rd}$ order, combinations of polynomial with dipole (pol+dipol, pol$\times$ dipol) with more than 4 free parameters, and rational forms with a variety of orders in (n,m) exhibit remarkable stability in their results as a function of the varying orders. For each group of functions the result is determined as the weighted average of the results from the individual fits, e.g. similar to the procedure followed in~\cite{Bernauer:2013tpr}. The process is repeated by varying the fitting range, with the $Q^2_{\rm max}$ taking values between $Q^2=0.25~({\rm GeV/c})^2$ and $Q^2=1~({\rm GeV/c})^2$, to explore their stability to the fitted range of 4-momentum transfer squared. The results for each one of the fitted functions are shown in Fig.~\ref{fig7rev} and for the different groups of functions are shown in Fig.~\ref{fig2}. A very good stability is observed in the final results for both the choice of the fitted functional forms and the dependence on the fitted range in momentum transfer as shown in Fig.~\ref{fig3}; above $Q^2=0.5~({\rm GeV/c})^2$ the results have also reached a high-level of statistical precision. 

\section{Results and discussion}
\begin{figure}
\centering
\resizebox{\linewidth}{!}{%
\includegraphics{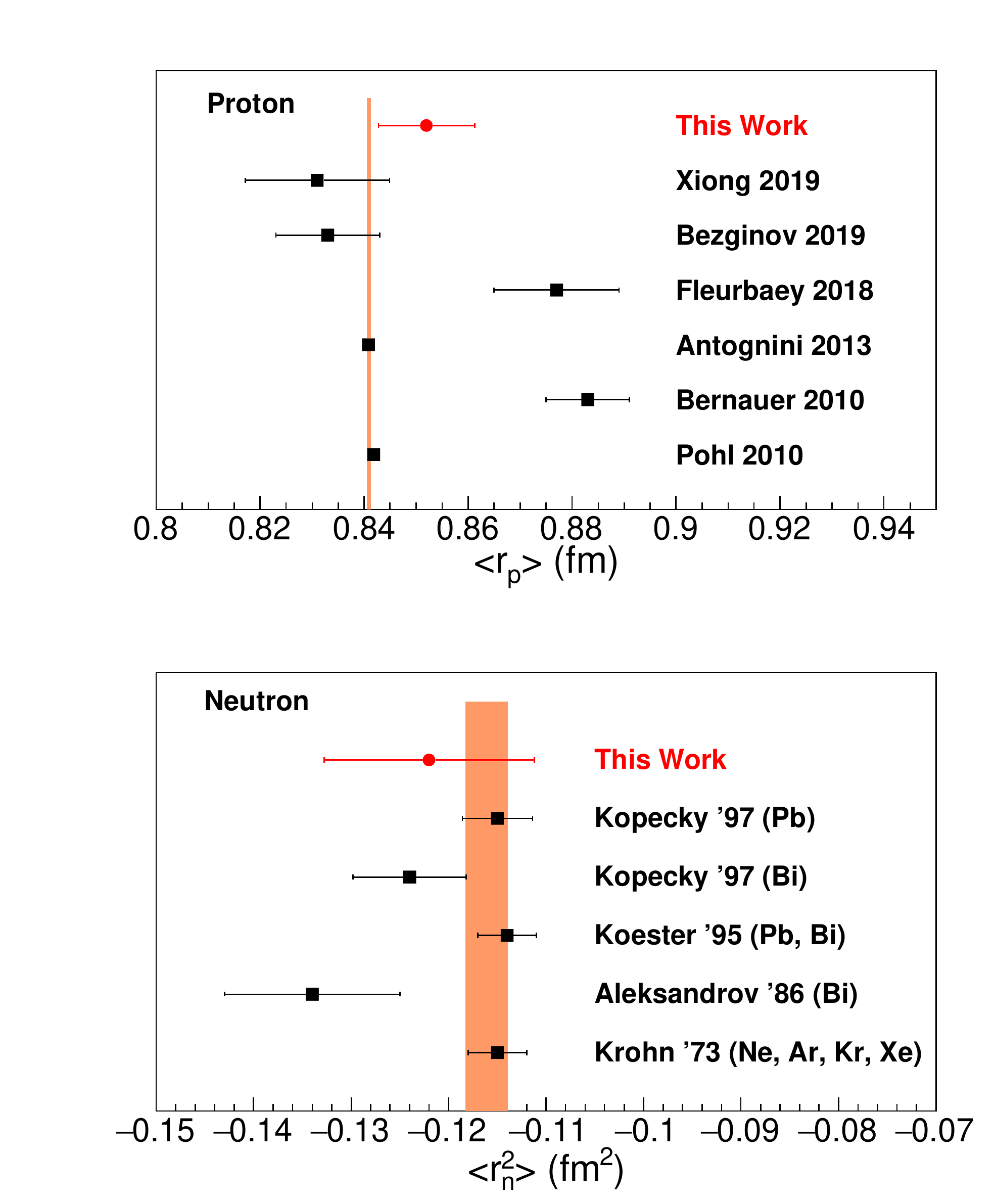}
}
\caption{Top panel: The proton charge radius extracted from  this work is shown along with the measurements of ~\cite{Bernauer.2010.105.242001,Xiong:2019umf,pohl:2010,bezginov:2019,Flaurbaey:2018,antog2013}. Bottom panel: The $\langle r_n^2 \rangle$ extracted  from this work is shown along with the neutron-electron scattering length measurements~\cite{Kopecky:1997rw,Koester:1995nx,Aleksandrov:1986mw,Krohn:1973re} that are currently included in the PDG $\langle r_n^2 \rangle$ analysis.}
\label{fig4}
\end{figure}

For the proton, all groups of considered functions are able to determine the charge radius with a very good precision. The polynomial and the polynomial+dipole forms exhibit a similar level of uncertainty when both statistical and systematic variations are considered, with the polynomials achieving a higher statistical precision. 
The rational forms follow, achieving a similar level of precision as the one achieved in~\cite{Xiong:2019umf}, in which only the (1,1) rational form~\cite{rat11} was used for extracting the radius.
For the neutron, all groups of functions agree nicely in the radius extraction, but it is only the polynomial and the polynomial+dipole forms that offer a good level of precision. For the neutron, all groups of functions are accounted for in the final extraction of the charge radius, but the effect of the polynomial$\times$dipole and of the rational forms is inconsequential, as a result of both their large uncertainties, as well as of the excellent agreement of their central values when compared to the results from the other two groups of functions (i.e. the polynomial and the polynomial+dipole). The final results for the proton and for the neutron charge radius are derived from the weighted average of all the fitted functions and are shown in Fig.~\ref{fig4}.

The systematic uncertainties include the dependence of the results on the selection of different functional forms adopted in the fits, as well as the stability of the results as a function of the fitted $Q^2$-range.  The systematic errors have been quantified from the weighted variance of the results for all individual fitted functional forms and for the different fitted ranges in momentum transfer. 
The uncertainties of the parametrizations  discussed in section~2 have been adopted in the flavor decomposition analysis, and they are thus subsequently accounted for in the extracted results for the charge radii. For completeness, the resulting systematic effect due to choice of parametrization was also studied.
The analysis is repeated with the different parametrizations discussed in section 2, and the variance of the results for the extracted charge radii is quantified as a systematic uncertainty. 
For the proton we find $\langle r_p \rangle = 0.852 \pm0.002_{\rm (stat.)} \pm0.009_{\rm (syst.)}~({\rm fm})$. Our result is consistent with the proton charge radius obtained from muonic hydrogen spectroscopy and from~\cite{Xiong:2019umf} and disagrees with the result of~\cite{Bernauer.2010.105.242001} (see Fig.~\ref{fig4}). Here we note that the dispersion theoretical analysis of the nucleon electromagnetic form factors has long pointed towards the smaller proton charge radius e.g. in~\cite{meis1,meis2,meis3,meis4}, while more recent efforts~\cite{doug,griff,mihov,alarcon,horbat} also point towards the same conclusion.

Recently, a precise determination of the proton charge radius was provided by the PRad collaboration~\cite{Xiong:2019umf}. PRad is the first experiment going down to $Q^2=2\times10^{-4}~(GeV/c)^2$ and makes use of a magnetic-spectrometer-free measurement off a windowless target which offers reduced systematic uncertainties. The PRad analysis derives $\langle r_p \rangle = 0.831 \pm0.007_{\rm (stat.)} \pm0.012_{\rm (syst.)}~({\rm fm})$, and utilizes only the PRad data which extend in momentum transfer up to $Q^2 \approx 0.06~({\rm GeV/c})^2$; no other world data are employed in the PRad fit. An interesting observation here is that, on their higher $Q^2$ end, the PRad measurements disagree with the data from~\cite{Bernauer.2010.105.242001}. Questions that naturally arise involve the reason for this discrepancy, and subsequently the effect on the radius extraction when a complete world data set is considered in the proton radius extraction. The latter question is addressed in this work where we have employed a complete data set for both nucleon form factors, that contains both the PRad measurements and previous world data.  Our result is found in good agreement with the PRad value~\cite{Xiong:2019umf}. Furthermore, an important conclusion is that the current method improves the precision of the $\langle r_p \rangle$ extraction from the form factor measurements, for both statistical and systematic uncertainties, as compared to the published results in~\cite{Xiong:2019umf}.

Another study was performed repeating our analysis while excluding the PRad data~\cite{Xiong:2019umf} from the data base.
In this case, the proton form factor data are primarily driven by the measurements described in~\cite{Bernauer.2010.105.242001,Bernauer:2013tpr,bern-thesis} which reported a large proton charge radius of $\approx 0.88~{\rm fm}$. 
However, in our analysis, excluding the PRad data and fitting ranges that extend above $Q^2=0.7~({\rm GeV/c})^2$, so that the convergence and the stability of the results is ensured, we derive again a small value for the proton charge radius, that is $ 0.857(13)~{\rm fm}$, but with a larger uncertainty. The fact that a smaller proton charge radius than~\cite{Bernauer.2010.105.242001,Bernauer:2013tpr} is derived without the inclusion of the PRad~\cite{Xiong:2019umf} data in the analysis indicates that the radius extraction method in~\cite{Bernauer.2010.105.242001,Bernauer:2013tpr} has most likely underestimated the level of the underlying uncertainties, or failed to avoid some form of bias in the fitting methodology. Thus, with this study we reach another conclusion, namely that we do not observe a discrepancy between the proton charge radius results derived from the electron scattering experiments~\cite{Bernauer.2010.105.242001} and \cite{Xiong:2019umf}, both of which are consistent with the proton charge radius obtained  from muonic hydrogen spectroscopy. We note that the experimental measurements in~\cite{Bernauer.2010.105.242001} are of very high quality and allow the reliable extraction of the proton charge radius when the methodology presented in this work is followed.

The current analysis offers the first extraction of the neutron charge radius utilizing the nucleon form factor data, and we find $\langle r_n^2 \rangle = -0.122 \pm0.004_{\rm (stat.)} \pm0.010_{\rm (syst.)}~({\rm fm}^2)$. The result agrees with the neutron-electron scattering length measurements, as shown in Fig.~\ref{fig4}.  The result is also in agreement with a recent neutron charge radius calculation that is based on the determination of the deuteron structure radius in chiral effective field theory and relying on atomic data for the difference of the deuteron and proton charge radii, giving $\langle r_n^2 \rangle =-0.105^{+0.005}_{-0.006}~({\rm fm}^2)$~\cite{bonnrn,filin2}. 
The neutron-electron scattering length measurements exhibit discrepancies with each other, displaying a $\approx 10\%$  tension between the results, suggesting that there are still unidentified systematic uncertainties associated with this method of extraction; as a consequence of this, the PDG~\cite{pdg} considers disagreeing measurements in the current evaluation of the $\langle r_n^2 \rangle$ world data average value. The precision of the new $\langle r_n^2 \rangle$ result is not sufficient in order to address these discrepancies. Nevertheless, the analysis presented here opens new possibilities to improve the precision of the $\langle r_n^2 \rangle$ extraction when the upcoming nucleon form factor measurements at the low~$Q^2$ frontier will become available in the near future. Forthcoming measurements focusing on the proton~\cite{protonrev} involve experiments at PSI (MUSE)~\cite{muse}, at Jefferson Lab (PRad-II), at MAMI and in Japan (ULQ$^2$). New experimental programs focusing on the neutron form factors at low momentum transfers are currently planned at Jefferson Lab~\cite{loi}, which can be further extended at MAMI/A1, as well as at high $Q^2$ utilizing the SBS experimental setup at Jefferson Lab.
Our studies have shown that with the inclusion of the upcoming measurements, this method of charge radius extraction will improve the precision of $\langle r_n^2 \rangle$ by more than a factor of two, thus making possible the resolution of the neutron-electron scattering length discrepancies via the comparison to an alternative method of extraction.


\section{Summary}
In summary, we have performed a simultaneous extraction of both the proton and the neutron charge radius from a combined analysis of proton and neutron form factor data, based on the flavor decomposition of the form factors. The method benefits the extraction of both nucleon's radii as it extends the data base by a collective analysis of both isospin partner form factor data.  Furthermore, it allows for a unified treatment in terms of the functional forms that are fitted for the radius extraction, bypassing the inherently different treatment of the proton and of the neutron electric form factor parametrizations. The first conclusion is that, within the context of the proton radius puzzle, we find a small value for the charge radius, but more importantly the new method improves the precision of the $\langle r_p \rangle$ extraction that is based on the form factor measurements. The second conclusion is that the current method overcomes biases of previous methods of the charge radius extraction and concludes that there is no true discrepancy between the proton charge radius results that have been derived by the electron scattering experiments, all of which are consistent with the proton charge radius obtained from muonic hydrogen spectroscopy. Another conclusion is that for the neutron we achieve the first extraction of its charge radius that is based on the only physical quantity that may be used to describe a formal neutron mean-square charge radius, namely the derivative of the electric form factor at $Q^2=0$.
The new method of extraction opens the path for a further improvement of the $\langle r_n^2 \rangle$ determination, and will offer a mutual benefit to the future extractions of both nucleon charge radii that will include upcoming form factor measurements.

\section*{Acknowledgments}
We would like to thank M.~Vanderhaeghen as this work received great benefit from his input and suggestions. This work has been supported by the US Department of Energy Office of Science, office of Nuclear Physics under contract no. DE-SC0016577, DE-FG02-94ER4084 and DEAC02-06CH11357. M.C. acknowledges financial support by the U.S. Department of Energy, Office of Nuclear Physics, Early Career Award under Grant No.\ DE-SC0020405.

%

\bibliographystyle{elsarticle-num}
\bibliography{V2EPJA}

\end{document}